\documentstyle[pra,aps]{revtex}

\begin{document}
\title{{\LARGE Generation of fan-states of radiation field in a cavity}}
\author{{\bf Nguyen Ba An}$^{a,b}$\thanks{%
Corresponding author. Email: nbaan@netnam.org.vn} and {\bf Truong Minh Duc}$%
^{c}$}
\address{$^{a}$Institute of Physics, P.O. Box 429 Bo Ho, Hanoi 10000, Vietnam\\
$^{b}$Faculty of Technology, Vietnam National University, 144 Xuan Thuy,\\
Cau Giay, Hanoi, Vietnam\\
$^{c}$Physics Department, Hue University, 32 Lo Loi, Hue, Vietnam}
\maketitle
\date{}

\begin{abstract}
A scheme of generating recently introduced fan-states $\left| \alpha
,2k\right\rangle _{F}$ ($\alpha $ is complex, $k=1,2,3,...$) is proposed
basing on a $\Lambda $-type atom-cavity field interaction. We show that with
suitable atomic preparations and measurements a passage of a sequence of $N$
atoms through a cavity may transform an initial field coherent state $\left|
\alpha \right\rangle $ to a fan-state $\left| \alpha ,2k\right\rangle _{F}$
with $k=2^{N-2}.$

PACS number(s): 42.50.Dv.
\end{abstract}

\pacs{42.50.Dv.}

\noindent {\bf 1. Introduction}

Nonclassical states are numerous though almost all of them are not {\it a
priori }available{\it \ }in the natural world but need be prepared by
appropriate experimental arrangements. Generation schemes prove very
important because without them a state could be looked upon as artificial
only. For example, the well-known squeezed state formulated theoretically in
1970 \cite{1} remained unphysical until 1985 when it was for the first time
observed in the laboratory by means of a four-wave mixing technique \cite{2}%
. Many nonclassical states are superposed of a finite or infinite, discrete
or continuous number of classical states. Superposition states may exhibit
various intriguing nonclassical properties such as squeezing, antibunching,
oscillatory number distribution, self-splitting, etc. thanks to quantum
interferences between their component states even though the latter are
purely classical. Of special interest are the Schr\"{o}dinger cats \cite{3}
which are composed of a pair of macroscopically distinguishable states. Best
documented types of Schr\"{o}dinger cats are states of the forms $\Psi
\propto \exp (-i\pi /4)\left| \alpha \right\rangle +\exp (i\pi /4)\left|
-\alpha \right\rangle $ (where $\left| \alpha \right\rangle $ with $\alpha
\in C$ is the coherent state of radiation field) which can be produced by
letting a coherent state propagate through an amplitude-dispersive medium 
\cite{4}, $\Psi \propto \left| \alpha \right\rangle \pm \left| -\alpha
\right\rangle $ which are called even/odd coherent state and can be produced
via a good deal of mechanisms (by quantum-nondemolition method \cite{5},
using a Mach-Zehnder interferometer \cite{6}, via a competitive two-photon
process \cite{7}, via Kerr nonlinearity \cite{8}, ...), $\Psi \propto \sqrt{%
\xi }\left| \alpha _{1}\right\rangle +\exp (i\phi )\sqrt{1-\xi }\left|
\alpha _{2}\right\rangle $ ($\xi \in [0,1],$ $\phi \in [0,2\pi ]$) which can
be produced by passing excited fast two-level atom through a very high-Q
cavity containing a field in the vacuum state, $\Psi \propto \left| \alpha
\right\rangle -\exp (-i\vartheta )\left| \alpha \exp (-i\vartheta
)\right\rangle $ which can be produced via the interaction of one atom with
a cavity supporting two different modes \cite{9}, etc.. During a last few
years Schr\"{o}dinger cats have been extended to the motion of a trapped ion 
\cite{10,11,12,12a,12b,13,14,15,16,17} as well as to atomic systems \cite
{18,19,20,21}. Various universal-like schemes have also been suggested for
preparing an arbitrary quantum superposition of coherent or Fock states
(see, e.g. \cite{22,22a,23,24,25,26,26a,26b}). Of particular interest is the
circle-state \cite{27,28,29,30,31,32} that consists of a discrete number of
coherent states equidistantly distributed on a circle in phase space.
Generation of circle-states in a trapped ion has been proposed recently in 
\cite{33}. Nonlinear coherent states \cite{33a} as new nonclassical states
have also been studied intensively during last years.

Recently, there has been introduced the so called fan-state \cite{34} whose
peculiarity is that in these states amplitude squeezing may arise
simultaneously in more than one direction leading to a multi-winged
flower-like (instead of an ellipse-like) shape of the uncertainty region.
The property of multi-directional squeezing would prove very useful in
future applications because information could be impressed on and extracted
from a number of amplitude components with decreased fluctuations. This
would provide potential parallel processing with high accuracy in
communication systems. However, the question of how to produce fan-states
for radiation field has not been touched upon so far. In the present paper
we consider a possible scheme to realize that.

\vskip 0.5cm

\noindent {\bf 2. Fan-states: definition}

Before defining the fan-state let us first mention what is called $K$-photon
coherent states \cite{35,36,36a,37,38,39}. These are the $K$ right
eigen-states, $\left| \alpha ,K,j\right\rangle $ with $\alpha \in C,$ $%
K=1,2,...,$ $j=0,1,...,K-1,$ of the operator $a^{K}$ ($a$ the photon
annihilation operator) 
\begin{equation}
a^{K}\left| \alpha ,K,j\right\rangle =\alpha ^{K}\left| \alpha
,K,j\right\rangle .
\end{equation}
As was proved in \cite{36a,37,38,39}, for given $K$ and $j,$ the state $%
\left| \alpha ,K,j\right\rangle $ can be expressed in terms of $K$ usual
coherent states as 
\begin{equation}
\left| \alpha ,K,j\right\rangle ={\cal N}(r^{2},K,j)\sum_{l=0}^{K-1}\text{e}%
^{2i\pi jl/K}\left| \alpha _{l}\right\rangle ,\text{ }\alpha _{l}=\alpha 
\text{e}^{2i\pi l/K}  \label{Kj}
\end{equation}
where ${\cal N}(r^{2},K,j)$ is the normalization factor with $r=|\alpha |.$
In the special case of $j=0$ we have from Eq. (\ref{Kj}) 
\begin{equation}
\left| \alpha ,K\right\rangle ={\cal N}(r^{2},K)\sum_{q=0}^{K-1}\left|
\alpha _{q}\right\rangle ,\text{ }\alpha _{q}=\alpha \text{e}^{2i\pi q/K}
\label{K}
\end{equation}
where we have identified $\left| \alpha ,K\right\rangle $ with $\left|
\alpha ,K,0\right\rangle $ and ${\cal N}(r^{2},K)$ with ${\cal N}%
(r^{2},K,0). $ Unlike states (\ref{Kj}), states (\ref{K}) have all the
components $\left| \alpha _{q}\right\rangle $ equally weighted. Thus they
are invariant with respect to the rotation $\alpha \rightarrow \alpha $e$%
^{2i\pi /K}$. With use of the scalar product 
\begin{equation}
\left\langle \beta \right. \left| \alpha \right\rangle =\exp \left[ \frac{1}{%
2}\left( 2\beta ^{*}\alpha -|\alpha |^{2}-|\beta |^{2}\right) \right]
\label{ab}
\end{equation}
of coherent states we are able to derive the normalization factor ${\cal N}%
(r^{2},K)$ for an arbitrary $K$ in the explicit form

\begin{equation}
{\cal N}(r^{2},K)=\left\{ K+2\sum_{q=1}^{K-1}q\cos \left[ r^{2}\sin (2\pi
q/K)\right] \exp \left[ r^{2}\left( \cos (2\pi q/K)-1\right) \right]
\right\} ^{-2}.  \label{Nc}
\end{equation}
It is easy to verify \cite{39} that, while the state $\left| \alpha
,K\right\rangle $ with $K$ odd is neither even nor odd (in the sense of a $%
\pi $-rotation $\alpha \rightarrow -\alpha ),$ it is even for $K$ even. We
shall be interested in even $K,$ i.e. $K=2k$ with $k=1,2,...$ and define the
fan-state $\left| \alpha ,2k\right\rangle _{F}$ characterized by $k$ in
terms of a linear superposition of states $\left| \alpha ,2k\right\rangle $
as 
\begin{equation}
\left| \alpha ,2k\right\rangle _{F}={\cal N}_{F}(r^{2},k)\sum_{p=0}^{2k-1}%
\left| \alpha _{p},2k\right\rangle ,\text{ }\alpha _{p}=\alpha \text{e}%
^{i\pi p/2k}  \label{fan}
\end{equation}
where ${\cal N}_{F}(r^{2},k)$, the normalization factor for the fan-state,
is found to be 
\begin{equation}
{\cal N}_{F}(r^{2},k)=\frac{{\cal N}(r^{2},4k)}{k{\cal N}(r^{2},2k)}
\label{M}
\end{equation}
with the ${\cal N}^{\prime }s$ given by Eq. (\ref{Nc}). The name ``fan''
comes from the orientation in complex plane of the radius-vectors $\alpha
_{p}=\alpha $e$^{i\pi p/2k}$ that looks like an open paper fan (see Fig. 2
in \cite{34}). In Figs. 1 and 2 we display the Q-function \cite{Q}
associated with the fan-state, 
\begin{equation}
Q_{k,\alpha }(\xi )=\frac{1}{\pi }\left| \left\langle \xi \right. \left|
\alpha ,2k\right\rangle _{F}\right| ^{2},  \label{Q}
\end{equation}
and its corresponding contour plot versus $x=\Re (\xi )$ and $y=\Im (\xi )$
for two sets of $k$ and $\alpha .$ As seen from these figures, there are
peaks and fringes between them. The fringes result from interferences
between the peaks and may trigger intriguing nonclassical effects such as
presence of simultaneous squeezing along $2k$ directions, as investigated in
detail in \cite{34}.

\vskip 0.5cm

\noindent {\bf 3. Hamiltonian and evolution operator}

In \cite{24} a method based on a Raman-type ($\Lambda $-configuration)
interaction between three-level atoms and radiation field in a lossless
cavity was outlined to produce quantum superpositions of coherent states of
the field. There was however a shortcoming in \cite{24} (see later) and the
treatment there was not delicate enough for a specific purpose. Here we
consider again the model in greater detail aiming specifically at preparing
the fan-state defined above. We start from the same effective interaction
Hamiltonian for the atom-field system as in \cite{24} 
\begin{equation}
H_{int}=-\lambda a^{+}aW  \label{1}
\end{equation}
where the operator $W$ is given by 
\begin{equation}
W=\left| +\right\rangle \left\langle -\right| +\left| -\right\rangle
\left\langle +\right| +\left| -\right\rangle \left\langle -\right| +\left|
+\right\rangle \left\langle +\right| .  \label{1a}
\end{equation}
In Eqs. (\ref{1}) and (\ref{1a}) $a$ ($a^{+}$) stands for the cavity field
annihilation (creation) operator, $\lambda $ for the effective coupling
constant and $\left| -\right\rangle $ ($\left| +\right\rangle )$ for the
ground (first excited) state of an atom (the second excited state $\left|
u\right\rangle $ of the atoms is inactive). Note that in deriving Eq. (\ref
{1}) the distance from the uppermost level $\left| u\right\rangle $ of the
three-level atom to its $\left| \pm \right\rangle $-levels was assumed well
detuned from the cavity mode to adiabatically eliminate $\left|
u\right\rangle $ and the matrix elements of the two transitions $\left|
u\right\rangle \leftrightarrow \left| \pm \right\rangle $ were set equal for
simplicity. It is easy to verify that 
\begin{equation}
W^{l}=2^{l-1}W
\end{equation}
for $l=1,2,3,...$and, hence, 
\begin{equation}
(H_{int})^{l}=\frac{1}{2}(-2\lambda a^{+}a)^{l}W  \label{2}
\end{equation}
leading to the evolution operator $U(t)$ of the form 
\begin{eqnarray}
U(t) &=&\exp \left( -iH_{int}t\right) =\sum_{l=0}^{\infty }\frac{\left(
-iH_{int}t\right) ^{l}}{l!}=1+\sum_{l=1}^{\infty }\frac{\left(
-iH_{int}t\right) ^{l}}{l!}  \nonumber \\
&=&1+\frac{1}{2}W\sum_{l=1}^{\infty }\frac{\left( 2i\lambda ta^{+}a\right)
^{l}}{l!}  \nonumber \\
&=&1+\frac{1}{2}W\left( \sum_{l=0}^{\infty }\frac{\left( 2i\lambda
ta^{+}a\right) ^{l}}{l!}-1\right)   \nonumber \\
&=&1+\frac{W}{2}\left( \text{e}^{2i\lambda ta^{+}a}-1\right) .  \label{UU}
\end{eqnarray}
Introducing the dimensionless time $\tau =2\lambda t$ gives 
\begin{equation}
U(\tau )=1+\frac{W}{2}\left( \text{e}^{i\tau a^{+}a}-1\right) .  \label{U}
\end{equation}

\vskip 0.5cm

\noindent {\bf 4. Fan-states: generation}

Consider a cavity containing initially at $\tau =\tau _{0}=0$ the radiation
field in a coherent state $\left| \alpha \right\rangle .$ We then send the
atoms one by one through the cavity. The $n^{th}$ ($n=1,2,...$) atom
prepared in an entangled state $\left| F_{n}\right\rangle =\xi _{n}\left|
-\right\rangle +\eta _{n}\left| +\right\rangle $ enters the cavity at time $%
\tau =\sum_{j=0}^{n-1}\tau _{j},$ spends a duration of $\tau _{n}$ in
interaction with the cavity field and is detected on its going out from the
cavity at time $\tau =\sum_{j=1}^{n}\tau _{j}$ in the state $\left|
S_{n}\right\rangle $ which is either $\left| -\right\rangle $ or $\left|
+\right\rangle .$ Within the model under consideration (the $\Lambda $-type
configuration) the uppermost level is always empty so that $\xi _{n}$ and $%
\eta _{n}$ should satisfy the condition $\left| \xi _{n}\right| ^{2}+\left|
\eta _{n}\right| ^{2}=1.$ The shortcoming in \cite{24} is that the authors
anticipated $\left| F_{n}\right\rangle =\varepsilon _{n}\left|
-\right\rangle +\left| +\right\rangle $ for which only $\varepsilon _{n}=0,$ 
$\forall n$ are allowed and therefore any argumentations made in \cite{24}
on a choice of $\varepsilon _{n}$ would be nonsensical ! The state of the
radiation field left inside the cavity after the detection of the $N^{th}$
atom is 
\begin{eqnarray}
\Phi _{\left| S_{1}\right\rangle ...\left| S_{N}\right\rangle }^{\left|
F_{1}\right\rangle ...\left| F_{N}\right\rangle }(\tau _{1}+...+\tau _{N})
&=&U_{\left| S_{N}\right\rangle }^{\left| F_{N}\right\rangle }(\tau
_{N})\Phi _{\left| S_{1}\right\rangle ...\left| S_{N-1}\right\rangle
}^{\left| F_{1}\right\rangle ...\left| F_{N-1}\right\rangle }(\tau
_{1}+...+\tau _{N-1})  \nonumber \\
&=&U_{\left| S_{N}\right\rangle }^{\left| F_{N}\right\rangle }(\tau
_{N})U_{\left| S_{N-1}\right\rangle }^{\left| F_{N-1}\right\rangle }(\tau
_{N-1})\Phi _{\left| S_{1}\right\rangle ...\left| S_{N-2}\right\rangle
}^{\left| F_{1}\right\rangle ...\left| F_{N-2}\right\rangle }(\tau
_{1}+...+\tau _{N-2})  \nonumber \\
&=&...  \nonumber \\
&=&U_{\left| S_{N}\right\rangle }^{\left| F_{N}\right\rangle }(\tau
_{N})U_{\left| S_{N-1}\right\rangle }^{\left| F_{N-1}\right\rangle }(\tau
_{N-1})...U_{\left| S_{3}\right\rangle }^{\left| F_{3}\right\rangle }(\tau
_{3})U_{\left| S_{2}\right\rangle }^{\left| F_{2}\right\rangle }(\tau
_{2})\Phi _{\left| S_{1}\right\rangle }^{\left| F_{1}\right\rangle }(\tau
_{1})  \nonumber \\
&=&U_{\left| S_{N}\right\rangle }^{\left| F_{N}\right\rangle }(\tau
_{N})U_{\left| S_{N-1}\right\rangle }^{\left| F_{N-1}\right\rangle }(\tau
_{N-1})...U_{\left| S_{2}\right\rangle }^{\left| F_{2}\right\rangle }(\tau
_{2})U_{\left| S_{1}\right\rangle }^{\left| F_{1}\right\rangle }(\tau
_{1})\left| \alpha \right\rangle  \label{PHI}
\end{eqnarray}
where 
\begin{equation}
U_{\left| S_{n}\right\rangle }^{\left| F_{n}\right\rangle }(\tau
_{n})=\left\langle s_{n}\right| U(\tau _{n})\left| F_{n}\right\rangle
\label{UFS}
\end{equation}
with $s_{n}=\pm 1$ if $S_{n}=\left| \pm \right\rangle $ and $U(\tau _{n})$
given by Eq. (\ref{U}). Making use of Eqs. (\ref{UFS}) and (\ref{U}) in Eq. (%
\ref{PHI}) yields 
\[
\Phi _{\left| S_{1}\right\rangle ...\left| S_{N}\right\rangle }^{\left|
F_{1}\right\rangle ...\left| F_{N}\right\rangle }(\tau _{1}+...+\tau _{N})=%
\frac{1}{2^{N}}\left\{ \prod_{j=1}^{N}s_{j}\left( \eta _{j}-\xi _{j}\right)
\left| \alpha \right\rangle \right. 
\]
\[
+\sum_{L=1}^{N-1}\left[ \sum_{l_{L}>...>l_{1}=1}^{N}\left( \prod_{p\neq
l_{1},...,l_{L}}^{N}s_{p}\left( \eta _{p}-\xi _{p}\right)
\prod_{q=l_{1}}^{l_{L}}s_{q}\left( \eta _{q}+\xi _{q}\right) \right) \left|
\alpha \text{e}^{i(\tau _{l_{1}}+...+\tau _{l_{L}})}\right\rangle \right] 
\]
\begin{equation}
\left. +\prod_{j=1}^{N}\left( \eta _{j}+\xi _{j}\right) \left| \alpha \text{e%
}^{i(\tau _{1}+...+\tau _{N})}\right\rangle \right\} .  \label{g}
\end{equation}
In general, for $\tau _{p}\neq \tau _{q},$ the state (\ref{g}) is linearly
superposed of $2^{N}$ distinguishable coherent states. The probability of
generating this superposition at time $\tau =\sum_{j=1}^{N}\tau _{j}$ is 
\begin{equation}
P_{\left| S_{1}\right\rangle ...\left| S_{N}\right\rangle }^{\left|
F_{1}\right\rangle ...\left| F_{N}\right\rangle }(\tau _{1}+...+\tau
_{N})=\left| \Phi _{\left| S_{1}\right\rangle }^{\left| F_{1}\right\rangle
}(\tau _{1})\Phi _{\left| S_{1}\right\rangle \left| S_{2}\right\rangle
}^{\left| F_{1}\right\rangle \left| F_{2}\right\rangle }(\tau _{1}+\tau
_{2})...\Phi _{\left| S_{1}\right\rangle ...\left| S_{N}\right\rangle
}^{\left| F_{1}\right\rangle ...\left| F_{N}\right\rangle }(\tau
_{1}+...+\tau _{N})\right| ^{2}
\end{equation}
because the success of having $\Phi _{\left| S_{1}\right\rangle ...\left|
S_{N}\right\rangle }^{\left| F_{1}\right\rangle ...\left| F_{N}\right\rangle
}$ at $\tau =\tau _{1}+...+\tau _{N}$ requires the success of having $\Phi
_{\left| S_{1}\right\rangle ...\left| S_{N-1}\right\rangle }^{\left|
F_{1}\right\rangle ...\left| F_{N-1}\right\rangle }$ at $\tau =\tau
_{1}+...+\tau _{N-1}$ which in turns requires the success of having $\Phi
_{\left| S_{1}\right\rangle ...\left| S_{N-2}\right\rangle }^{\left|
F_{1}\right\rangle ...\left| F_{N-2}\right\rangle }$ at $\tau =\tau
_{1}+...+\tau _{N-2}$ and so on. Specially, if each of the atoms is prepared
at its entrance and detected at its exit in the same state, i.e. $\left|
F_{j}\right\rangle \equiv \left| S_{j}\right\rangle ,$ then (\ref{g})
simplifies to 
\begin{equation}
\Phi _{\left| S_{1}\right\rangle ...\left| S_{N}\right\rangle }^{\left|
S_{1}\right\rangle ...\left| S_{N}\right\rangle }(\tau _{1}+...+\tau _{N})=%
\frac{1}{2^{N}}\left\{ \left| \alpha \right\rangle +\sum_{L=1}^{N-1}\left[
\sum_{l_{L}>...>l_{1}=1}^{N}\left| \alpha \text{e}^{i(\tau _{l_{1}}+...+\tau
_{l_{L}})}\right\rangle \right] +\left| \alpha \text{e}^{i(\tau
_{1}+...+\tau _{N})}\right\rangle \right\}  \label{11}
\end{equation}
in which all the components are equally weighted. Next, if the velocities of
the atoms are selected such that $\tau _{j}=\pi /2^{j-1}$ then at time $\tau
=\sum_{j=1}^{N}\tau _{j}=(2-2^{1-N})\pi $ the state (\ref{11}) becomes 
\begin{equation}
\Phi _{\left| S_{1}\right\rangle ...\left| S_{N}\right\rangle }^{\left|
S_{1}\right\rangle ...\left| S_{N}\right\rangle }((2-2^{1-N})\pi )=\frac{%
\left| \alpha ,2^{N-1}\right\rangle _{F}}{2^{N}{\cal N}(r^{2},2^{N})}
\label{12}
\end{equation}
implying that the fan-state characterized by $k=2^{N-2}$ can be produced by
sending $N=2+\log _{2}(k)$ atoms (properly prepared, velocity-selected and
detected as described above) through the cavity. The total probability of
obtaining such fan-states is 
\begin{equation}
P_{2^{N-2}}=\prod_{n=1}^{N}\frac{1}{2^{2n}{\cal N}(r^{2},2^{n})}.  \label{14}
\end{equation}

For clarity we now demonstrate the proposed scheme for generating the
fan-state with $k=1.$ Obviously, the number of atoms to be sent is $N=2.$
The initial state of the atom-field system is factorized as $\Psi (0)=\left|
F_{1}\right\rangle \otimes \left| \alpha \right\rangle .$ Due to the
atom-field interaction inside the cavity, at time $\tau _{1}$ the atom-field
system becomes entangled whose state is described by 
\begin{eqnarray}
\Psi (\tau _{1}) &=&U(\tau _{1})\left| F_{1}\right\rangle \otimes \left|
\alpha \right\rangle  \nonumber \\
&=&\frac{1}{2}\left\{ \left[ \left( \xi _{1}-\eta _{1}\right) \left| \alpha
\right\rangle +\left( \xi _{1}+\eta _{1}\right) \left| \alpha \text{e}%
^{i\tau _{1}}\right\rangle \right] \left| -\right\rangle \right.  \nonumber
\\
&&+\left. \left[ \left( \eta _{1}-\xi _{1}\right) \left| \alpha
\right\rangle +\left( \xi _{1}+\eta _{1}\right) \left| \alpha \text{e}%
^{i\tau _{1}}\right\rangle \right] \left| +\right\rangle \right\}  \label{6}
\end{eqnarray}
which projects the field on the state 
\begin{equation}
\Phi _{\left| -\right\rangle }^{\left| F_{1}\right\rangle }(\tau _{1})=\frac{%
1}{2}\left[ \left( \xi _{1}-\eta _{1}\right) \left| \alpha \right\rangle
+\left( \xi _{1}+\eta _{1}\right) \left| \alpha \text{e}^{i\tau
_{1}}\right\rangle \right]  \label{6a}
\end{equation}
if the atom is detected in the ground state $\left| S_{1}\right\rangle
=\left| -\right\rangle ,$ or 
\begin{equation}
\Phi _{\left| +\right\rangle }^{\left| F_{1}\right\rangle }(\tau _{1})=\frac{%
1}{2}\left[ \left( \eta _{1}-\xi _{1}\right) \left| \alpha \right\rangle
+\left( \xi _{1}+\eta _{1}\right) \left| \alpha \text{e}^{i\tau
_{1}}\right\rangle \right]  \label{6b}
\end{equation}
if the atom is detected in the excited state $\left| S_{1}\right\rangle
=\left| +\right\rangle .$ The two formulae (\ref{6a}) and (\ref{6b}) can be
unified into a single one as 
\begin{equation}
\Phi _{\left| S_{1}\right\rangle }^{\left| F_{1}\right\rangle }(\tau _{1})=%
\frac{1}{2}\left[ s_{1}\left( \eta _{1}-\xi _{1}\right) \left| \alpha
\right\rangle +\left( \eta _{1}+\xi _{1}\right) \left| \alpha \text{e}%
^{i\tau _{1}}\right\rangle \right] =U_{\left| S_{1}\right\rangle }^{\left|
F_{1}\right\rangle }(\tau _{1})\left| \alpha \right\rangle  \label{7}
\end{equation}
with $U_{\left| S_{1}\right\rangle }^{\left| F_{1}\right\rangle }$ defined
by Eq. (\ref{UFS}). The probability of finding states (\ref{7}) at $\tau
_{1} $ is 
\[
P_{\left| S_{1}\right\rangle }^{\left| F_{1}\right\rangle }(\tau _{1})= 
\]
\begin{equation}
\frac{1}{2}\left\{ 1+s_{1}\text{e}^{-r^{2}(1-\cos \tau _{1})}\left[ \left(
|\eta _{1}|^{2}-|\xi _{1}|^{2}\right) \cos \left( r^{2}\sin \tau _{1}\right)
-2\Im \left( \xi _{1}^{*}\eta _{1}\right) \sin \left( r^{2}\sin \tau
_{1}\right) \right] \right\} .
\end{equation}
States (\ref{7}) generally represent superpositions of two arbitrary
coherent states whose weights are controlled by $\xi _{1},$ $\eta _{1}$ and
whose relative phases by the interaction duration $\tau _{1}.$ If we prepare
the atom initially in the ground state $\left| F_{1}\right\rangle =\left|
-\right\rangle ,$ i.e. $\eta _{1}=0$ ($\xi _{1}=1$), we success in having
the field states 
\begin{equation}
\Phi _{\left| -\right\rangle }^{\left| -\right\rangle }(\tau _{1})=\frac{1}{2%
}\left( \left| \alpha \right\rangle +\left| \alpha \text{e}^{i\tau
_{1}}\right\rangle \right) ,  \label{7a}
\end{equation}
\begin{equation}
\Phi _{\left| +\right\rangle }^{\left| -\right\rangle }(\tau _{1})=\frac{1}{2%
}\left( -\left| \alpha \right\rangle +\left| \alpha \text{e}^{i\tau
_{1}}\right\rangle \right)  \label{7b}
\end{equation}
with the probabilities 
\begin{equation}
P_{\left| \pm \right\rangle }^{\left| -\right\rangle }(\tau _{1})=\frac{1}{2}%
\left[ 1\mp \text{e}^{-r^{2}(1-\cos \tau _{1})}\cos \left( r^{2}\sin \tau
_{1}\right) \right] .  \label{7ab}
\end{equation}
Alternatively, if we prepare the atom initially in the excited state $\left|
F_{1}\right\rangle =\left| +\right\rangle ,$ i.e. $\eta _{1}=1$ ($\xi _{1}=0$%
), we success in having the field states

\begin{equation}
\Phi _{\left| -\right\rangle }^{\left| +\right\rangle }(\tau _{1})=\frac{1}{2%
}\left( -\left| \alpha \right\rangle +\left| \alpha \text{e}^{i\tau
_{1}}\right\rangle \right) ,  \label{7c}
\end{equation}
\begin{equation}
\Phi _{\left| +\right\rangle }^{\left| +\right\rangle }(\tau _{1})=\frac{1}{2%
}\left( \left| \alpha \right\rangle +\left| \alpha \text{e}^{i\tau
_{1}}\right\rangle \right) .  \label{7d}
\end{equation}
with the probabilities 
\begin{equation}
P_{\left| \pm \right\rangle }^{\left| +\right\rangle }(\tau _{1})=\frac{1}{2}%
\left[ 1\pm \text{e}^{-r^{2}(1-\cos \tau _{1})}\cos \left( r^{2}\sin \tau
_{1}\right) \right] .  \label{7cd}
\end{equation}
From Eqs. (\ref{7ab}) and (\ref{7cd}) we realize that 
\begin{equation}
P_{\left| +\right\rangle }^{\left| +\right\rangle }(\tau _{1})=P_{\left|
-\right\rangle }^{\left| -\right\rangle }(\tau _{1})\text{ and }P_{\left|
-\right\rangle }^{\left| +\right\rangle }(\tau _{1})=P_{\left|
+\right\rangle }^{\left| -\right\rangle }(\tau _{1}).
\end{equation}
Figure 4 plots $P_{\left| \pm \right\rangle }^{\left| \pm \right\rangle }$
and $P_{\left| \mp \right\rangle }^{\left| \pm \right\rangle }$ versus $\tau
_{1}$ for several values of $r.$ Transparently, at $\tau _{1}=\pi $ (or $%
3\pi ,$ $5\pi ,...),$ the states $\Phi _{\left| \pm \right\rangle }^{\left|
\pm \right\rangle }$ ($\Phi _{\left| \mp \right\rangle }^{\left| \pm
\right\rangle })$ are proportional to an even (odd) coherent state. Even
coherent states are necessary precursors for generating fan-sates in next
steps.

To generate fan-states with $k=1$ sending a second atom at time $\tau _{1}$
is needed. Let the initial state of the second atom be $\left|
F_{2}\right\rangle .$ After an interaction duration $\tau _{2},$ when the
second atom is detected at its exit in state $\left| S_{2}\right\rangle ,$
the field will be projected on the state 
\begin{eqnarray}
\Phi _{\left| S_{1}\right\rangle \left| S_{2}\right\rangle }^{\left|
F_{1}\right\rangle \left| F_{2}\right\rangle }(\tau _{1}+\tau _{2}) &=&\frac{%
1}{4}\left[ s_{1}s_{2}\left( \eta _{1}-\xi _{1}\right) \left( \eta _{2}-\xi
_{2}\right) \left| \alpha \right\rangle +s_{2}\left( \eta _{1}+\xi
_{1}\right) \left( \eta _{2}-\xi _{2}\right) \left| \alpha \text{e}^{i\tau
_{1}}\right\rangle \right.  \nonumber \\
&&\left. +s_{1}\left( \eta _{1}-\xi _{1}\right) \left( \eta _{2}+\xi
_{2}\right) \left| \alpha \text{e}^{i\tau _{2}}\right\rangle +\left( \eta
_{1}+\xi _{1}\right) \left( \eta _{2}+\xi _{2}\right) \left| \alpha \text{e}%
^{i(\tau _{1}+\tau _{2})}\right\rangle \right]  \label{8}
\end{eqnarray}
which can be formulated in the form 
\begin{equation}
\Phi _{\left| S_{1}\right\rangle \left| S_{2}\right\rangle }^{\left|
F_{1}\right\rangle \left| F_{2}\right\rangle }(\tau _{1}+\tau
_{2})=U_{\left| S_{2}\right\rangle }^{\left| F_{2}\right\rangle }(\tau
_{2})\Phi _{\left| S_{1}\right\rangle }^{\left| F_{1}\right\rangle }(\tau
_{1}).
\end{equation}
The final probability of having $\Phi _{\left| S_{1}\right\rangle \left|
S_{2}\right\rangle }^{\left| F_{1}\right\rangle \left| F_{2}\right\rangle
}(\tau _{1}+\tau _{2})$ is 
\begin{equation}
P_{\left| S_{1}\right\rangle \left| S_{2}\right\rangle }^{\left|
F_{1}\right\rangle \left| F_{2}\right\rangle }(\tau _{1}+\tau _{2})=\left|
\Phi _{\left| S_{1}\right\rangle }^{\left| F_{1}\right\rangle }(\tau
_{1})\Phi _{\left| S_{1}\right\rangle \left| S_{2}\right\rangle }^{\left|
F_{1}\right\rangle \left| F_{2}\right\rangle }(\tau _{1}+\tau _{2})\right|
^{2}.
\end{equation}
For $\tau _{1}=\tau _{2}$ the state (\ref{8}) reduces to a superposition of
three coherent states whose weights are always different. Nevertheless, for $%
\tau _{1}\neq \tau _{2},$ it is superposed of four coherent states whose
weights may be made equal by suitably preparing the atomic states at the
entrance to the cavity. Four particular situations are of interest.

i) Both the first and second atoms are prepared at their entrance in the
excited state $\left| F_{1}\right\rangle =\left| F_{2}\right\rangle =\left|
+\right\rangle $, i.e. $\eta _{1}=\eta _{2}=1$ $(\xi _{1}=\xi _{2}=0),$
leading to 
\begin{equation}
\Phi _{\left| +\right\rangle \left| +\right\rangle }^{\left| +\right\rangle
\left| +\right\rangle }(\tau _{1}+\tau _{2})=\frac{1}{4}\left( \left| \alpha
\right\rangle +\left| \alpha \text{e}^{i\tau _{1}}\right\rangle +\left|
\alpha \text{e}^{i\tau _{2}}\right\rangle +\left| \alpha \text{e}^{i(\tau
_{1}+\tau _{2})}\right\rangle \right) ,  \label{9a}
\end{equation}
\begin{equation}
\Phi _{\left| +\right\rangle \left| -\right\rangle }^{\left| +\right\rangle
\left| +\right\rangle }(\tau _{1}+\tau _{2})=\frac{1}{4}\left( -\left|
\alpha \right\rangle -\left| \alpha \text{e}^{i\tau _{1}}\right\rangle
+\left| \alpha \text{e}^{i\tau _{2}}\right\rangle +\left| \alpha \text{e}%
^{i(\tau _{1}+\tau _{2})}\right\rangle \right) ,  \label{9a1}
\end{equation}
\begin{equation}
\Phi _{\left| -\right\rangle \left| +\right\rangle }^{\left| +\right\rangle
\left| +\right\rangle }(\tau _{1}+\tau _{2})=\frac{1}{4}\left( -\left|
\alpha \right\rangle +\left| \alpha \text{e}^{i\tau _{1}}\right\rangle
-\left| \alpha \text{e}^{i\tau _{2}}\right\rangle +\left| \alpha \text{e}%
^{i(\tau _{1}+\tau _{2})}\right\rangle \right) ,  \label{9a2}
\end{equation}
\begin{equation}
\Phi _{\left| -\right\rangle \left| -\right\rangle }^{\left| +\right\rangle
\left| +\right\rangle }(\tau _{1}+\tau _{2})=\frac{1}{4}\left( \left| \alpha
\right\rangle -\left| \alpha \text{e}^{i\tau _{1}}\right\rangle -\left|
\alpha \text{e}^{i\tau _{2}}\right\rangle +\left| \alpha \text{e}^{i(\tau
_{1}+\tau _{2})}\right\rangle \right) .  \label{9a3}
\end{equation}

ii) The first (second) atom is prepared at its entrance in its excited $%
\left| F_{1}\right\rangle =\left| +\right\rangle $, i.e. $\eta _{1}=1$ $(\xi
_{1}=0)$ (ground $\left| F_{2}\right\rangle =\left| -\right\rangle $, i.e. $%
\eta _{2}=0$ $(\xi _{2}=1))$ state, leading to 
\begin{equation}
\Phi _{\left| +\right\rangle \left| +\right\rangle }^{\left| +\right\rangle
\left| -\right\rangle }(\tau _{1}+\tau _{2})=\frac{1}{4}\left( -\left|
\alpha \right\rangle -\left| \alpha \text{e}^{i\tau _{1}}\right\rangle
+\left| \alpha \text{e}^{i\tau _{2}}\right\rangle +\left| \alpha \text{e}%
^{i(\tau _{1}+\tau _{2})}\right\rangle \right) ,  \label{9b}
\end{equation}
\begin{equation}
\Phi _{\left| +\right\rangle \left| -\right\rangle }^{\left| +\right\rangle
\left| -\right\rangle }(\tau _{1}+\tau _{2})=\frac{1}{4}\left( \left| \alpha
\right\rangle +\left| \alpha \text{e}^{i\tau _{1}}\right\rangle +\left|
\alpha \text{e}^{i\tau _{2}}\right\rangle +\left| \alpha \text{e}^{i(\tau
_{1}+\tau _{2})}\right\rangle \right) ,  \label{9b1}
\end{equation}
\begin{equation}
\Phi _{\left| -\right\rangle \left| +\right\rangle }^{\left| +\right\rangle
\left| -\right\rangle }(\tau _{1}+\tau _{2})=\frac{1}{4}\left( \left| \alpha
\right\rangle -\left| \alpha \text{e}^{i\tau _{1}}\right\rangle -\left|
\alpha \text{e}^{i\tau _{2}}\right\rangle +\left| \alpha \text{e}^{i(\tau
_{1}+\tau _{2})}\right\rangle \right) ,  \label{9b2}
\end{equation}
\begin{equation}
\Phi _{\left| -\right\rangle \left| -\right\rangle }^{\left| +\right\rangle
\left| -\right\rangle }(\tau _{1}+\tau _{2})=\frac{1}{4}\left( -\left|
\alpha \right\rangle +\left| \alpha \text{e}^{i\tau _{1}}\right\rangle
-\left| \alpha \text{e}^{i\tau _{2}}\right\rangle +\left| \alpha \text{e}%
^{i(\tau _{1}+\tau _{2})}\right\rangle \right) .  \label{9b3}
\end{equation}

iii) The first (second) atom is prepared at its entrance in its ground $%
\left| F_{1}\right\rangle =\left| -\right\rangle $, i.e. $\eta _{1}=0$ $(\xi
_{1}=1)$ (excited $\left| F_{2}\right\rangle =\left| +\right\rangle $, i.e. $%
\eta _{2}=1$ $(\xi _{2}=0)$) state, leading to 
\begin{equation}
\Phi _{\left| +\right\rangle \left| +\right\rangle }^{\left| -\right\rangle
\left| +\right\rangle }(\tau _{1}+\tau _{2})=\frac{1}{4}\left( -\left|
\alpha \right\rangle +\left| \alpha \text{e}^{i\tau _{1}}\right\rangle
-\left| \alpha \text{e}^{i\tau _{2}}\right\rangle +\left| \alpha \text{e}%
^{i(\tau _{1}+\tau _{2})}\right\rangle \right) ,  \label{9c}
\end{equation}
\begin{equation}
\Phi _{\left| +\right\rangle \left| -\right\rangle }^{\left| -\right\rangle
\left| +\right\rangle }(\tau _{1}+\tau _{2})=\frac{1}{4}\left( \left| \alpha
\right\rangle -\left| \alpha \text{e}^{i\tau _{1}}\right\rangle -\left|
\alpha \text{e}^{i\tau _{2}}\right\rangle +\left| \alpha \text{e}^{i(\tau
_{1}+\tau _{2})}\right\rangle \right) ,  \label{9c1}
\end{equation}
\begin{equation}
\Phi _{\left| -\right\rangle \left| +\right\rangle }^{\left| -\right\rangle
\left| +\right\rangle }(\tau _{1}+\tau _{2})=\frac{1}{4}\left( \left| \alpha
\right\rangle +\left| \alpha \text{e}^{i\tau _{1}}\right\rangle +\left|
\alpha \text{e}^{i\tau _{2}}\right\rangle +\left| \alpha \text{e}^{i(\tau
_{1}+\tau _{2})}\right\rangle \right) ,  \label{9c2}
\end{equation}
\begin{equation}
\Phi _{\left| -\right\rangle \left| -\right\rangle }^{\left| -\right\rangle
\left| +\right\rangle }(\tau _{1}+\tau _{2})=\frac{1}{4}\left( -\left|
\alpha \right\rangle -\left| \alpha \text{e}^{i\tau _{1}}\right\rangle
+\left| \alpha \text{e}^{i\tau _{2}}\right\rangle +\left| \alpha \text{e}%
^{i(\tau _{1}+\tau _{2})}\right\rangle \right) .  \label{9c3}
\end{equation}

iv) Both the first and second atoms are prepared at their entrance in their
ground state $\left| F_{1}\right\rangle =\left| F_{2}\right\rangle =\left|
-\right\rangle $, i.e. $\eta _{1}=\eta _{2}=0$ $(\xi _{1}=\xi _{2}=1),$
leading to 
\begin{equation}
\Phi _{\left| +\right\rangle \left| +\right\rangle }^{\left| -\right\rangle
\left| -\right\rangle }(\tau _{1}+\tau _{2})=\frac{1}{4}\left( \left| \alpha
\right\rangle -\left| \alpha \text{e}^{i\tau _{1}}\right\rangle -\left|
\alpha \text{e}^{i\tau _{2}}\right\rangle +\left| \alpha \text{e}^{i(\tau
_{1}+\tau _{2})}\right\rangle \right) ,  \label{9d}
\end{equation}
\begin{equation}
\Phi _{\left| +\right\rangle \left| -\right\rangle }^{\left| -\right\rangle
\left| -\right\rangle }(\tau _{1}+\tau _{2})=\frac{1}{4}\left( -\left|
\alpha \right\rangle +\left| \alpha \text{e}^{i\tau _{1}}\right\rangle
-\left| \alpha \text{e}^{i\tau _{2}}\right\rangle +\left| \alpha \text{e}%
^{i(\tau _{1}+\tau _{2})}\right\rangle \right) ,  \label{9d1}
\end{equation}
\begin{equation}
\Phi _{\left| -\right\rangle \left| +\right\rangle }^{\left| -\right\rangle
\left| -\right\rangle }(\tau _{1}+\tau _{2})=\frac{1}{4}\left( -\left|
\alpha \right\rangle -\left| \alpha \text{e}^{i\tau _{1}}\right\rangle
+\left| \alpha \text{e}^{i\tau _{2}}\right\rangle +\left| \alpha \text{e}%
^{i(\tau _{1}+\tau _{2})}\right\rangle \right) ,  \label{9d2}
\end{equation}
\begin{equation}
\Phi _{\left| -\right\rangle \left| -\right\rangle }^{\left| -\right\rangle
\left| -\right\rangle }(\tau _{1}+\tau _{2})=\frac{1}{4}\left( \left| \alpha
\right\rangle +\left| \alpha \text{e}^{i\tau _{1}}\right\rangle +\left|
\alpha \text{e}^{i\tau _{2}}\right\rangle +\left| \alpha \text{e}^{i(\tau
_{1}+\tau _{2})}\right\rangle \right) .  \label{9d3}
\end{equation}

It follows directly from above that the four component coherent states
contribute equally in four situations implied by 
\begin{equation}
\Phi _{\left| +\right\rangle \left| +\right\rangle }^{\left| +\right\rangle
\left| +\right\rangle }=\Phi _{\left| +\right\rangle \left| -\right\rangle
}^{\left| +\right\rangle \left| -\right\rangle }=\Phi _{\left|
-\right\rangle \left| +\right\rangle }^{\left| -\right\rangle \left|
+\right\rangle }=\Phi _{\left| -\right\rangle \left| -\right\rangle
}^{\left| -\right\rangle \left| -\right\rangle }.  \label{FS}
\end{equation}
These mean that if each of the two atoms is prepared at its entrance and
detected at its exit in the same pure state, i.e. $\left| F_{j}\right\rangle
=\left| S_{j}\right\rangle ,$ then the field reduces to a superposition of
four equally weighted coherent states, i.e. 
\begin{equation}
\Phi _{\left| S_{1}\right\rangle \left| S_{2}\right\rangle }^{\left|
S_{1}\right\rangle \left| S_{2}\right\rangle }(\tau _{1}+\tau _{2})=\frac{1}{%
4}\left( \left| \alpha \right\rangle +\left| \alpha \text{e}^{i\tau
_{1}}\right\rangle +\left| \alpha \text{e}^{i\tau _{2}}\right\rangle +\left|
\alpha \text{e}^{i(\tau _{1}+\tau _{2})}\right\rangle \right) ,  \label{SS}
\end{equation}
no matter $\left| S_{1,2}\right\rangle $ is $\left| -\right\rangle $ or $%
\left| +\right\rangle .$ For atomic velocity selections such that $\tau
_{1}=\pi $ $(\pi /2)$ and $\tau _{2}=\pi /2$ $(\pi )$ the superposition (\ref
{SS}) at $\tau =\tau _{1}+\tau _{2}=3\pi /2$ becomes 
\begin{equation}
\Phi _{\left| S_{1}\right\rangle \left| S_{2}\right\rangle }^{\left|
S_{1}\right\rangle \left| S_{2}\right\rangle }(3\pi /2)=\frac{1}{4}\left[
\left| \alpha \right\rangle +\left| -\alpha \right\rangle +\left| i\alpha
\right\rangle +\left| -i\alpha \right\rangle \right] =\frac{\left| \alpha
,2\right\rangle _{F}}{4{\cal N}(r^{2},4)}  \label{10}
\end{equation}
which is proportional to the simplest fan-state characterized by $k=1.$ The
final probability $P_{1}$ of observing the $k=1$ fan-state depends on $r,$ 
\begin{equation}
P_{1}(r)=\frac{1}{8}\left( 1+\text{e}^{-2r^{2}}+2\text{e}^{-r^{2}}\cos
r^{2}\right) \times \left\{ 
\begin{array}{lll}
\left( 1+\text{e}^{-2r^{2}}\right) & \text{for} & \tau _{1}=\pi ,\tau
_{2}=\pi /2 \\ 
\left( 1+\text{e}^{-r^{2}}\cos r^{2}\right) & \text{for} & \tau _{1}=\pi
/2,\tau _{2}=\pi
\end{array}
\right. ,  \label{P1}
\end{equation}
as depicted in Fig. 5.

Similarly, we can send a third atom to generate the $k=2$ fan-state with a
certain probability. In this way we find out that fan-states with $k=2^{N-2}$
($N=2,$ $3,$ $4,...$ is the number of atoms to be sent) can be generated
provided the atoms be appropriately prepared, velocity-selected and detected
as described above. Figure 6 illustrates the $r$-dependent probability of
finding the fan-state $\left| \alpha ,2k\right\rangle _{F}$ for several
values of $k$ at time $\tau =(2-2^{1-N})\pi $ when the duration the $j^{th}$
atom spends inside the cavity is controlled so that $\tau _{j}=\pi /2^{j-1}.$
This figure shows that the greater the value of $k$ the smaller the
generation probability and a finite probability maintains for all $k$ in the
small-$r$ domain where field amplitude squeezing is favorable as revealed in 
\cite{34}.

\vskip 0.5cm

\noindent {\bf 5. Conclusion}

To sum up we have developed further the method proposed in \cite{24} to
generate the fan-state in which multi-directional higher-order amplitude
squeezing is possible. We have shown that by sending a sequence of $N\geq 2$
atoms through a cavity initially containing radiation field in a coherent
state we are able to generate fan-states $\left| \alpha ,2k\right\rangle _{F}
$ with $k=2^{N-2}.$ For the success each of the atoms should enter and go
out from the cavity in the same state (either $\left| -\right\rangle $ or $%
\left| +\right\rangle )$. Moreover, the atoms should be velocity-selected so
that the $n^{th}$ atom spends the duration of $\tau _{n}=\pi /2^{n-1}$ in
interaction with the cavity field. Our analysis has also revealed that the
probability of success decreases with increasing $k$ and $|\alpha |.$ Of
interest has been the fact that the fan-state generation probability is
finite for small values of $|\alpha |$ for which amplitude squeezing always
exists \cite{34}.

\vskip 0.5cm

\noindent {\bf Acknowledgments}

This work was in part supported by the National Basic Research Program
KT-04.1.2 and the Faculty of Technology of VNU.

\begin{center}
\newpage

{\bf Figure captions}
\end{center}

\begin{enumerate}
\item[Fig. 1:]  Distribution function $\pi Q_{k,\alpha }(\xi )$ versus $%
x=\Re (\xi )$ and $y=\Im (\xi )$ for $k=1$ and $|\alpha |=2$ (left) and its
corresponding contour plot (right).

\item[Fig. 2:]  As in Fig. 1 but for $k=2$ and $|\alpha |=3.5.$

\item[Fig. 3:]  The time-dependent probability $P\equiv P_{\left| \pm
\right\rangle }^{\left| \pm \right\rangle }$ (dashed curves) and $P_{\left|
\mp \right\rangle }^{\left| \pm \right\rangle }$ (solid curves) for $r=0.5$
(top), $r=1.0$ (middle) and $r=5.0$ (bottom). Here ``Time'' denotes the
dimensionless time $\tau _{1}$ defined in the text.

\item[Fig. 4:]  The $r$-dependent probability $P_{1}$ of obtaining the $k=1$
fan-state for $\tau _{1}=\pi ,$ $\tau _{2}=\pi /2$ (solid curve) and $\tau
_{1}=\pi /2,$ $\tau _{2}=\pi $ (dashed curve).

\item[Fig. 5:]  The $r$-dependent probability $P_{k}$ of obtaining the
fan-state characterized by $k=2^{N-2}$ for $\tau _{j}=\pi /2^{j-1}$ with $%
j=1,$ $2,...,$ $N.$ The used $k$-values ($N$-values) are $k=1,$ $2,$ $4$ and 
$8$ ($N=2,$ $3,$ $4$ and $5)$ downwards.
\end{enumerate}

\end{document}